\documentclass[12pt]{iopart}
\usepackage{iopams}

\usepackage{epsfig}
\usepackage{setstack}
\usepackage{color}
\usepackage{url}
\usepackage{hyperref}

\begin{document}

\title{High Fidelity Detection of the Orbital Angular Momentum of Light by Time Mapping}
\author{Paul Bierdz, Minho Kwon, Connor Roncaioli, Hui Deng}
\address{Department of Physics, University of Michigan, Ann Arbor, MI 48109, USA}
\eads{\mailto{dengh@umich.edu}}
\begin{abstract} 
We demonstrate high-fidelity detection of the orbital angular momentum (OAM) of light using a compact and practical OAM spectrometer that maps the OAM spectrum to time. The spectrometer consists of a single optical delay loop to achieve timing mapping, a vortex phase plate that iteratively decreases the OAM value, and a single mode fibre to distinguish zero from non-zero OAM states. Light with arbitrarily OAM compositions can be measured. For light with OAM up to~$4\hbar$, we measured an average crosstalk of~-21.3 dB, which is mainly limited by the purity of the input states and optical alignment.
\end{abstract}
\maketitle

\section{Introduction}

The orbital angular momentum (OAM) of light has been exploited for a wide range of modern applications.  With a discrete, yet unbounded Hilbert space \cite{allen_orbital_1992}, OAM has been used to increase the capacity of free-space communications \cite{gibson_free-space_2004,wang_terabit_2012}, generate high order entanglement \cite{mair_entanglement_2001,dada_experimental_2011} and increase security of non-classical communications  \cite{durt_security_2004}.  The unique topological properties of OAM states have been utilized to create optical tweezers that can apply varying degrees of torque \cite{he_direct_1995,grier_revolution_2003}, high-sensitivity vortex coronagraphs for extrasolar planet detection\cite{foo_optical_2005} and revealing topological properties of objects\cite{torner_digital_2005}. A recent paper suggests that light emitted nearby rotating black holes is twisted such that the angular momentum of the black hole can be inferred from the OAM spectrum \cite{tamburini_twisting_2011}.

For these applications, it is often crucial to discern different OAM states with high fidelity.  Several methods have been demonstrated for measuring the OAM spectrum including those using fork diffraction gratings coupled with single mode fibres (SMF) \cite{mair_entanglement_2001}, cascading Mach-Zehnder interferometers \cite{leach_measuring_2002} and transformation optics \cite{berkhout_efficient_2010}.  However, all these methods map different OAM states to spatially separated modes; as a result, complexity and size of the experimental setups increase with the highest measurable OAM state.  Fidelity of the detection is often limited by the complexity of the experimental setup and optical diffraction. There are other techniques that do not spatially separate the modes.  Courtial et. al. proposed a scheme based on OAM-dependent rotational Doppler shift \cite{courtial_measurement_1998}; yet experimental implementation has not been possible for optical wavelengths. Recently, we proposed a scheme based on OAM to time mapping \cite{bierdz_compact_2011}. The scheme utilizes the concept of quantum counterfactual measurement and quantum Zeno effect to allow non-destructive sorting of the OAM spectrum with unit efficiency, if optical loss or misalignment is neglected.  Experimental implementation with lossy optics, however, will significantly reduce the efficiency for high-order OAM state.

Here we demonstrate a simplified, and practical, OAM-to-time mapping scheme, which achieved a record high extinction ratio among five OAM states at an operation speed of~80~MHz.  Since OAM is mapped to time, the same apparatus can be used to measure an arbitrary number of unique OAM states.  We note that a similar scheme was also adopted in a recent experiment on time-division multiplexing of OAM, although the fidelity and repetition rate were orders of magnitude lower than reported here \cite{karimi_time-division_2012}.  We also note that, without employing a quantum Zeno investigator, time-division schemes are not suitable for improving channel capacity in communication.
  
\section{Principle of the OAM spectrometer}
   
\begin{figure}[b]
\begin{center}\includegraphics[width=0.6\textwidth]{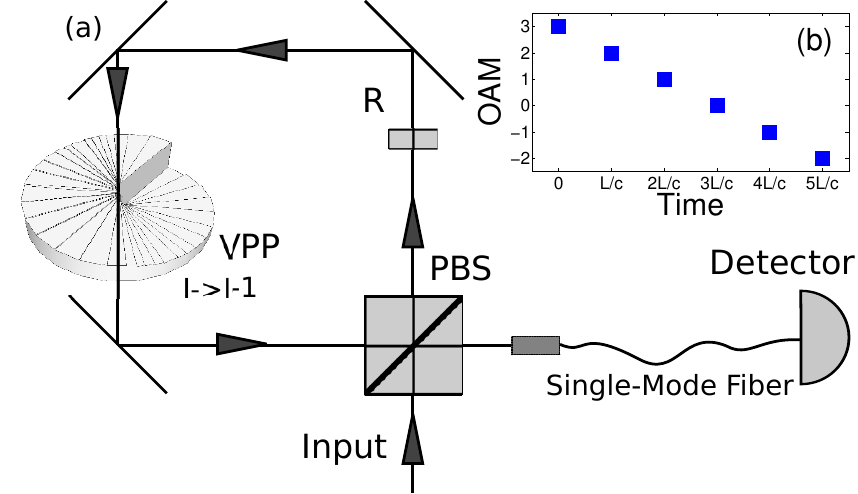}\end{center}
\caption{(a) Schematic of the OAM spectrometer, consisting of an optical loop that converts an input pulse into a sequence of pulses equally spaced in time, a VPP that decreases the OAM value by 1 per pass, and an SMF to filter out states with non-zero OAM. (b) Showing how the OAM value changes in time for an example input state with $\ell_0=3$.}
\label{fig:setup}
\end{figure}

As shown in figure~\ref{fig:setup}a, our OAM spectrometer consists of a single optical loop to perform time mapping, a vortex phase plate (VPP) of topological charge 1 as an OAM ladder operator, and an SMF as a filter for the fundamental Gaussian mode with zero OAM.  Consider an incident pulse consisting of a fraction $\beta_{\ell}$ of OAM components with OAM value $\ell$, where $\ell$ is an integer and $\sum_{\ell=-\infty}^\infty \beta_{\ell}=1$.  The optical loop converts the pulse into a sequence of pulses equally separated by the round trip propagation time $T$ in the loop.  The loop needs an even number of reflections to maintain the same sign of OAM.  Per loop, the VPP decreases the OAM value of each OAM component by $1$.  Hence, after $N$ loops, the fractional $\beta_{\ell}$ of the original pulse will have OAM value $\ell-N$.  Only the component with zero OAM value in each pulse can pass through the SMF to be detected.  Thus, the OAM component $\beta_{\ell}$ with OAM value $\ell$ in the original pulse will exit the spectrometer at a time $t=lT$ (figure~\ref{fig:setup}b).

\subsection{Energy distribution}

The distribution of the input-pulse energy among the sequence of pulses can be pre-calibrated and controlled by the beam splitter with polarization optics.  We control the distribution with a polarizing beam-splitter (PBS) and half-wave plate (labelled as R in figure~\ref{fig:setup}a) in the loop.  The incident light is first set to be linearly polarized at an angle $\phi_0$ with respect to the vertical.  The PBS splits off the fraction of $\sin^2\!\phi_0$ into the first time window and sends the rest, now vertically polarized, into the loop.  The wave plate R, rotated at an angle $\theta/2$ with respect to the vertical, rotates the vertically polarized light by an angle $\theta$.   The corresponding energy distribution is described below: 

\begin{eqnarray}
P(l)= \cases{
  \sin^2\!\phi_0                                         &  \text{if $\ell = 0$,} \\
  \cos^2\!\phi_0 \cos^2\!\theta                          &  \text{if $\ell = 1$,} \\
  \cos^2\!\phi_0 \sin^4\!\theta  \cos^{2(\ell-2)}\!\theta   &  \text{if $\ell \geq 2$.}}
\label{eq:p_pbs}
\end{eqnarray}

Here $\ell$ represents the $\ell$th output pulse.  Alternatively, the PBS and wave plate can be replaced by a non-polarizing beam-splitter with a chosen splitting ratio for polarization-insensitive measurements.  This would allow for information to be encoded in the polarization degree of freedom.
    
\section{Experimental implementation}
\label{sec:results}

We implemented an OAM spectrometer as illustrated in figure~\ref{fig:setup}(a). To test its performance, we used OAM eigenstates as input pulses and detected the output pulses using a Hamamatsu streak camera. 

The input pulse was generated by diffracting a pulsed Gaussian laser beam off fork-diffraction patterns \cite{bazhekov_laser-beams_1990} on a Holoeye PLUTO LCoS spatial light modulator (SLM) with a period of about~10 lines per millimeter and pixel pitch of~8$\mu$m.  The initial laser beam was from a Tsunami Ti-Sapphire laser centered at 730~nm, with a pulse width of~100~fs and repetition rate of~80~MHz, and collimated with a~10x objective lens from an SMF ensuring an initial $M^2$ close to~1 and a spot size within the frame of the SLM.  

\begin{figure}[h]
\includegraphics[width=1\textwidth]{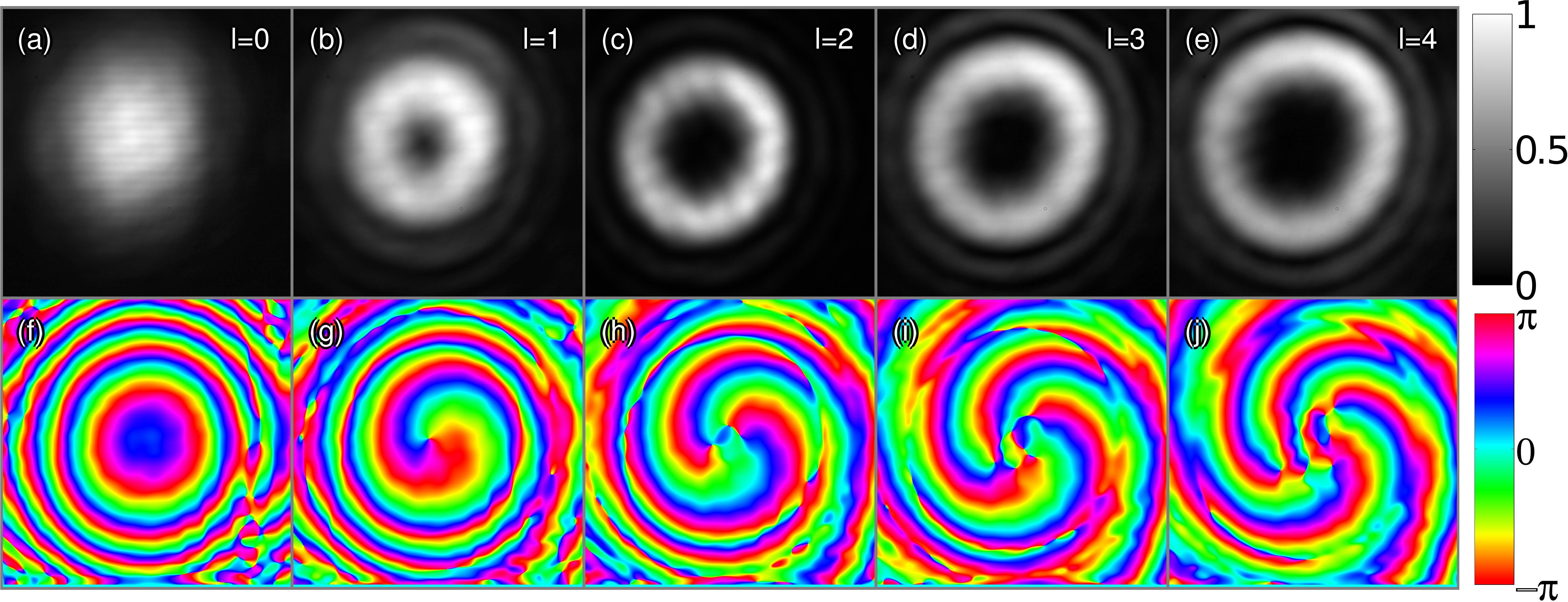}
\caption{(a-e) The intensity measured with a camera of the different initial OAM beams from the SLM with $\ell=0-4$.  (f-j) The phase fronts of the input OAM beams.  They are calculated from the interference patterns between the OAM beams and a reference Gaussian beam, as explained in the text.}
\label{fig:oam_beams}
\end{figure}

We first verify the input states by measuring its far-field intensity and phase distributions, as shown in figure~\ref{fig:oam_beams}.  The intensity distributions were measured with a charge-coupled device. The higher-order OAM beams show a larger spatial size as expected.  Inhomogeneity in the radial intensity distribution reflects the lack of purity of the OAM value.  

We measured the phase distribution by interfering the OAM beam with a reference Gaussian beam from the original laser.  This produces a forked interference pattern.  From this interference pattern we extract the phase front, as plotted in figure~\ref{fig:oam_beams}(f-j), by performing a Fourier transform, selecting the first diffracted mode, centring and then applying an inverse Fourier transform.  The phase should twist about by $2\pi\ell$ for an $\ell^{\mbox{th}}$-order OAM beam.  Deviation from this results implies impure initial OAM states from optical aberrations or phase ripples from modulating the liquid crystals.

To calibrate the time-mapping and energy distribution of the OAM spectrometer, we used the zero-OAM Gaussian state as the input and replaced the VPP by a flat glass plate of the same thickness.  The output from the SMF was detected by a Hamamatsu streak camera with a time resolution of~0.02~ns. As shown in figure~\ref{fig:setup}(b), the input pulse was mapped into a sequence of pulses separated by~$T=1.03$~ns.  The timing and dynamic range of the streak camera allowed us to measure the first five output pulses, or, the first five OAM states, at the laser repetition rate of 80 MHz.  The experiment can also be redesigned to allow the measurement of a larger number of pulses by using different laser repetition rates, loop sizes or measurement devices.

\begin{figure}
\includegraphics[width=0.5\textwidth]{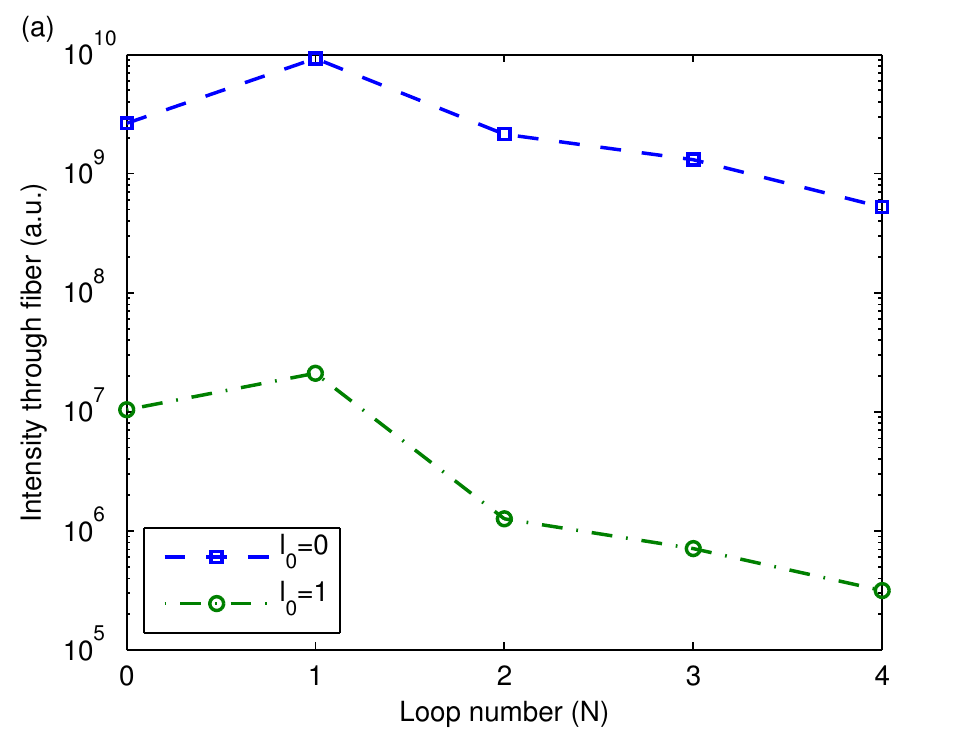}
\includegraphics[width=0.5\textwidth]{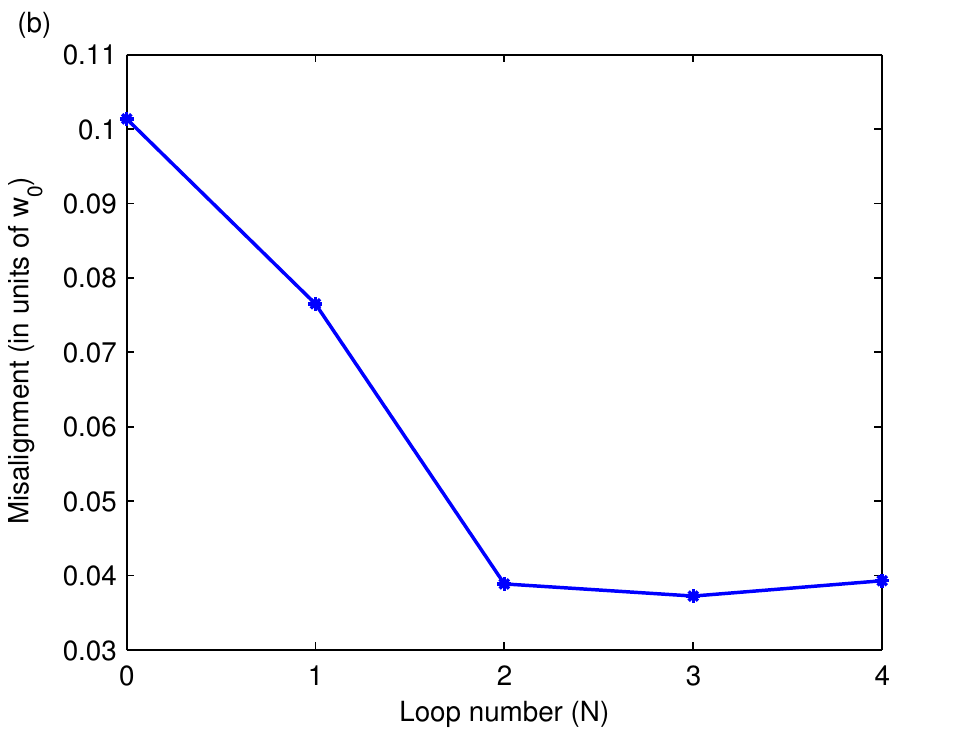}
\caption{(a) Unnormalized power through the fibre versus the number of passes through the loop without a VPP and different initial OAM values ($\ell_0=0,1$) set by the SLM.  (b) Computed misalignment of SMF.}
\label{fig:results}
\end{figure}

The energy distribution among the pulses was calculated by integrating the intensity over each output pulse. The results are shown in figure~\ref{fig:results}(a) (the $\ell_0=0$ curve).  We performed the same measurement with the input state $\ell_0=1$.  The ratio of the two energy normalization curves at each $\ell T$ gives an estimate of the misalignment of the SMF-coupling (figure~\ref{fig:results}(b)) for the $\ell^{\mbox{th}}$ output pulse, as will be discussed in detail in Section~\ref{sec:SMF_misalignment} later.  The uniformity of ratio among different $\ell$ confirms that the loop was well aligned. 

\begin{figure}
\includegraphics[width=0.33\textwidth]{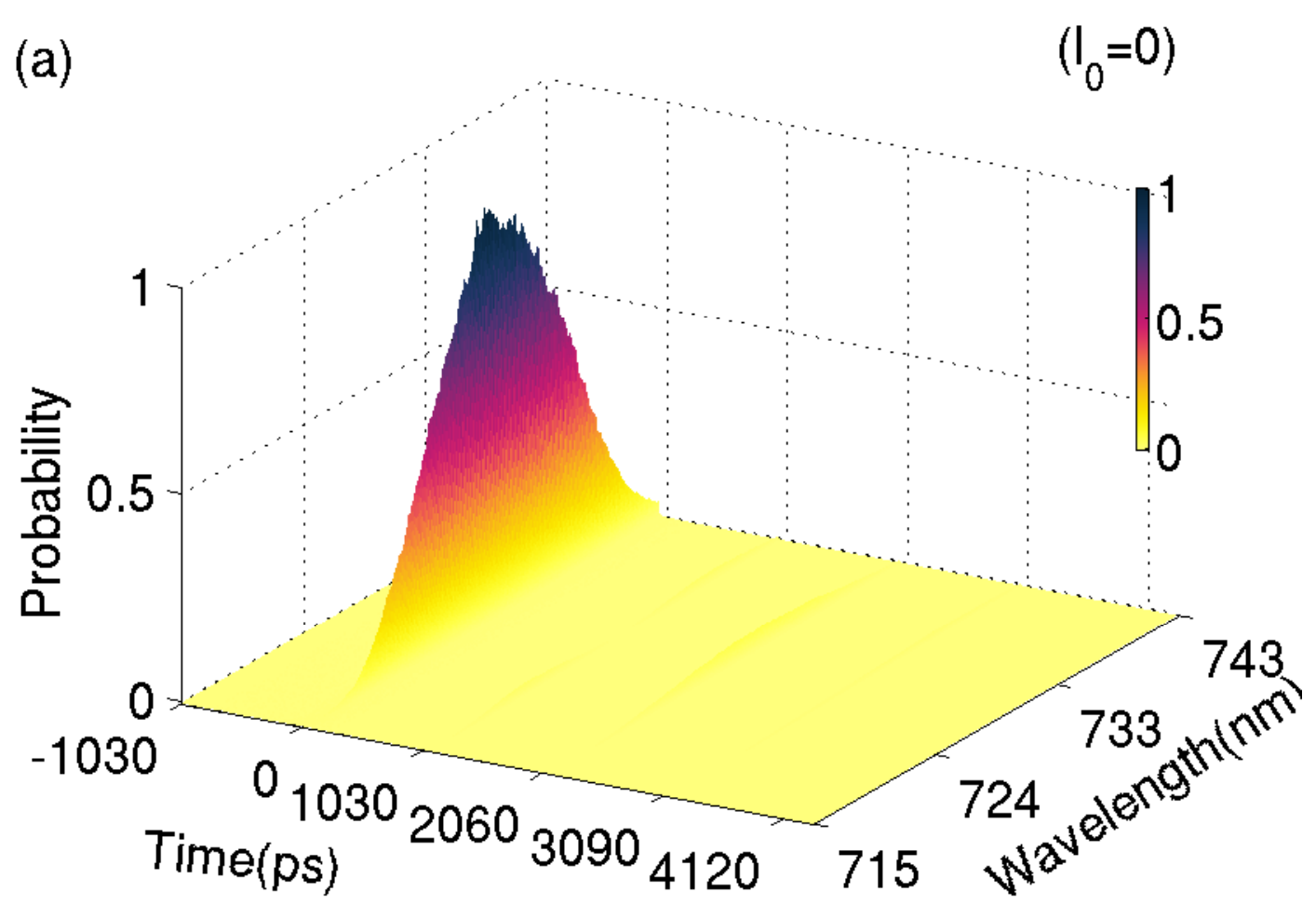}
\includegraphics[width=0.33\textwidth]{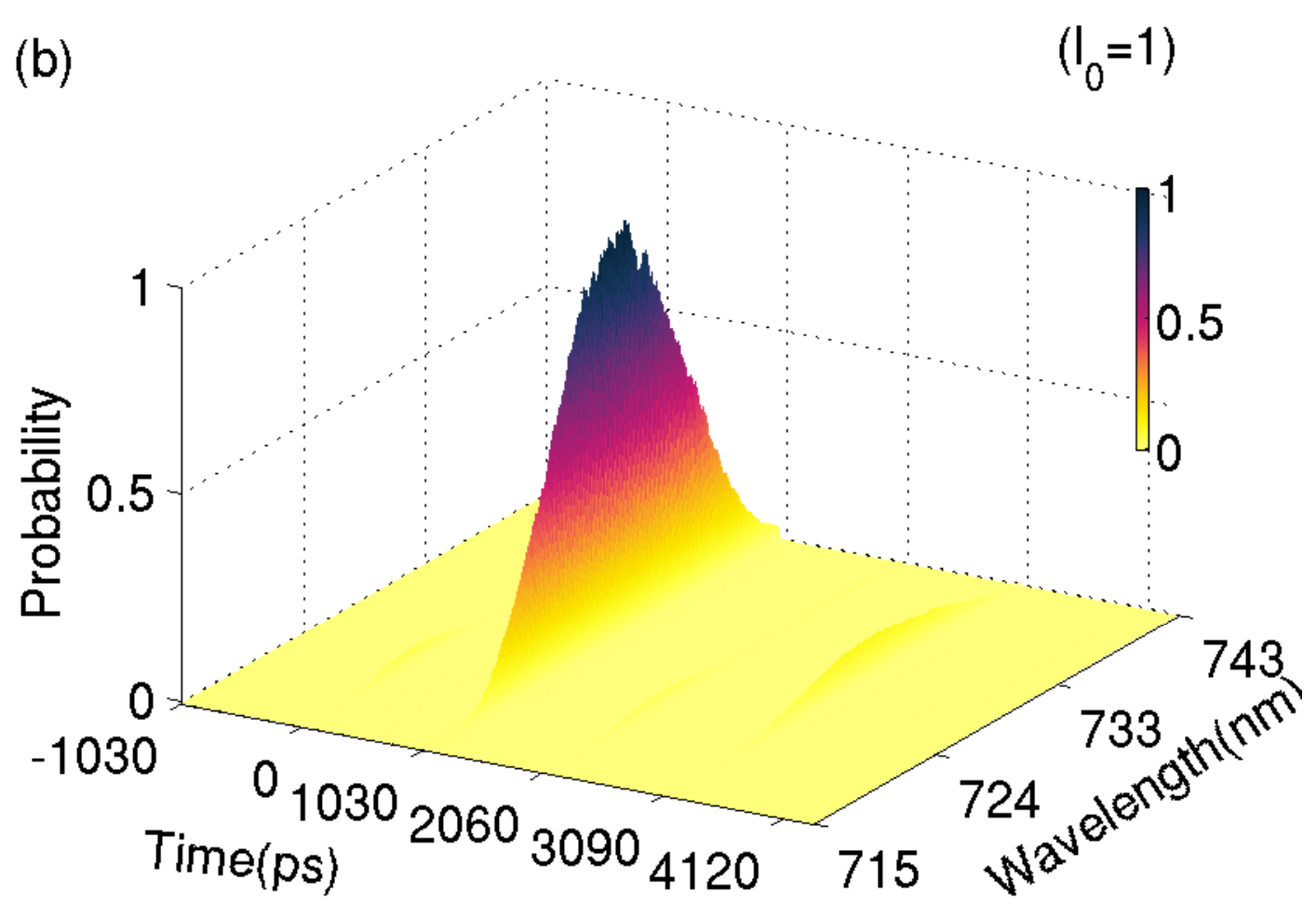}
\includegraphics[width=0.33\textwidth]{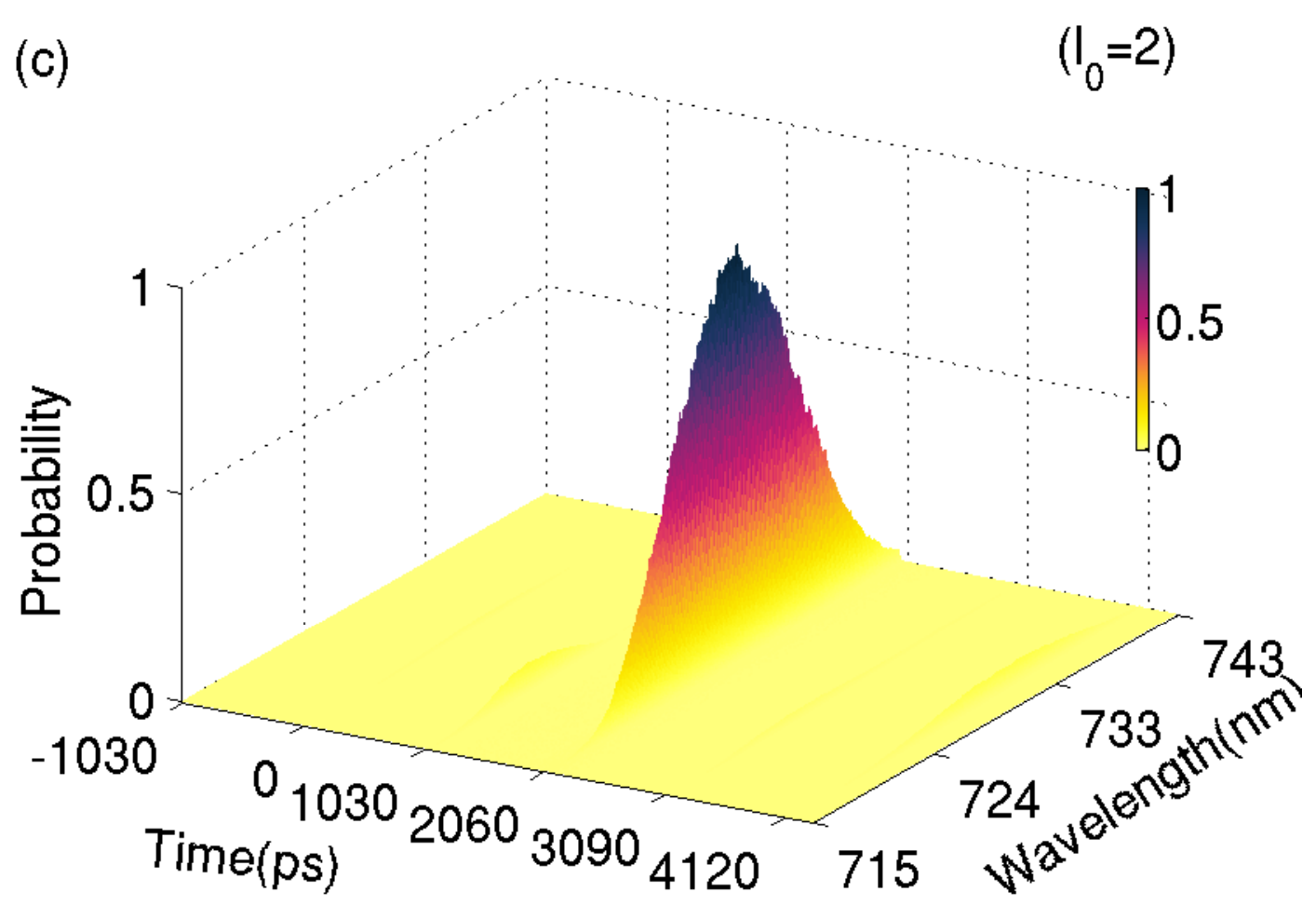}

\includegraphics[width=0.33\textwidth]{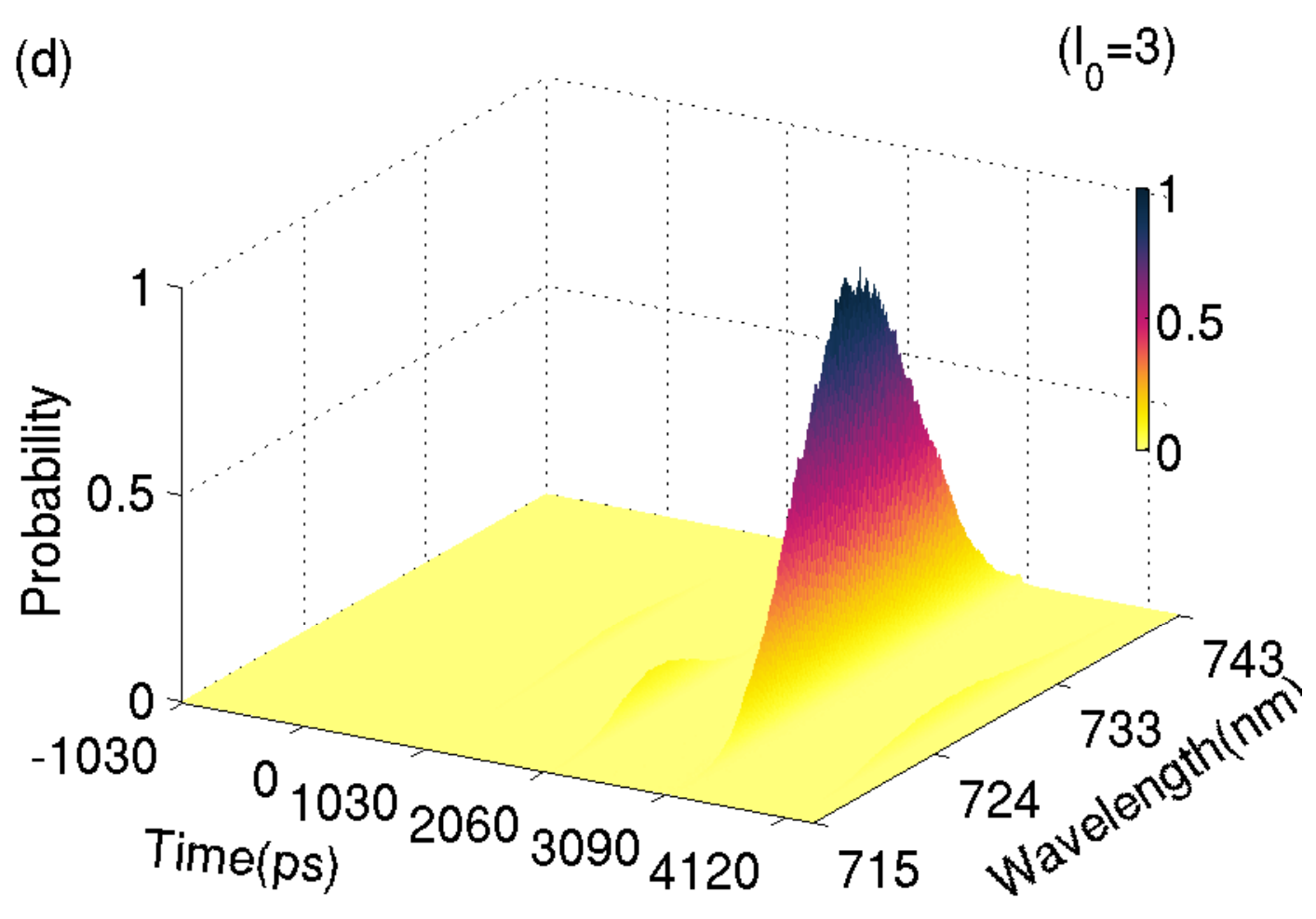}
\includegraphics[width=0.33\textwidth]{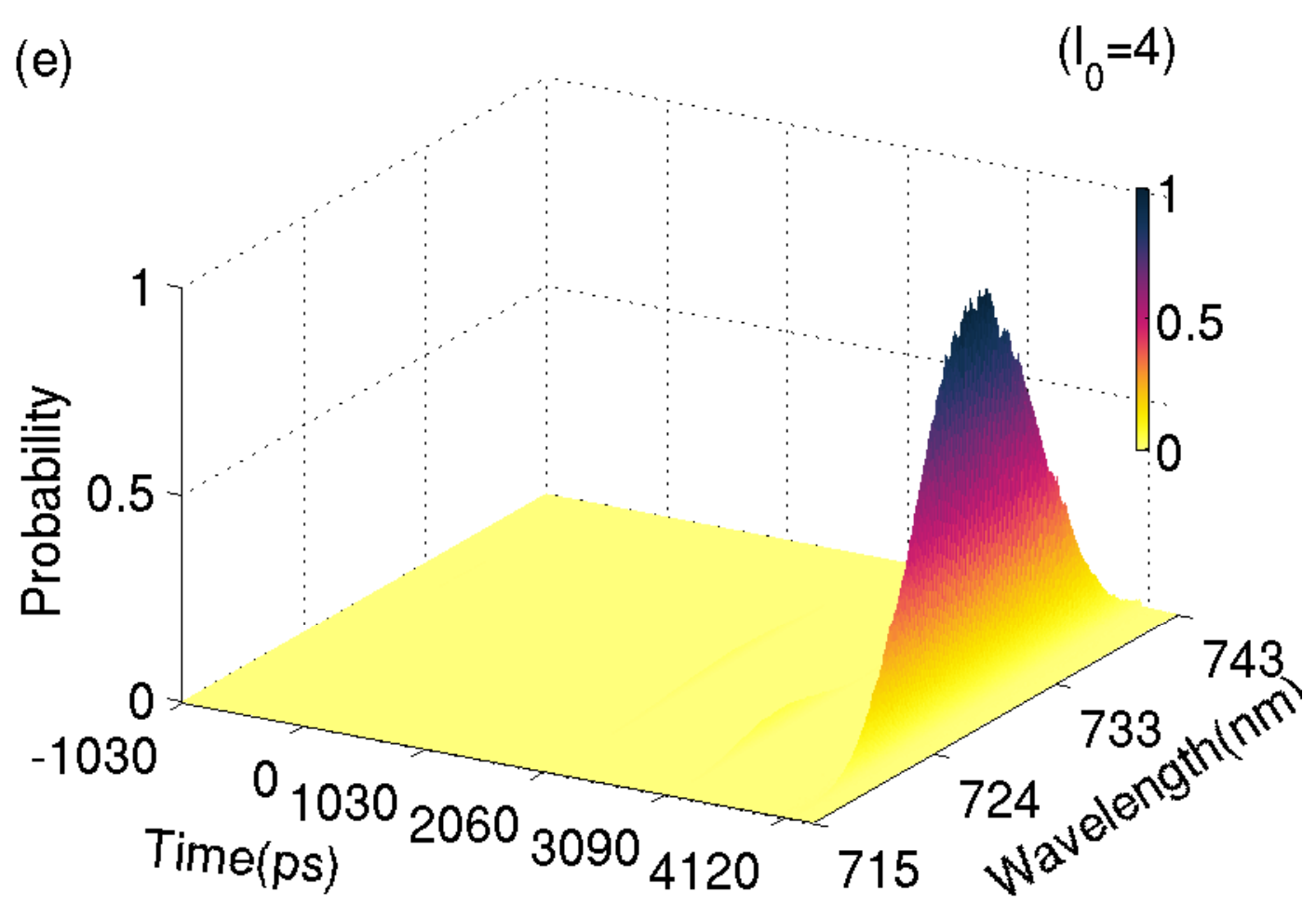}
\includegraphics[width=0.33\textwidth]{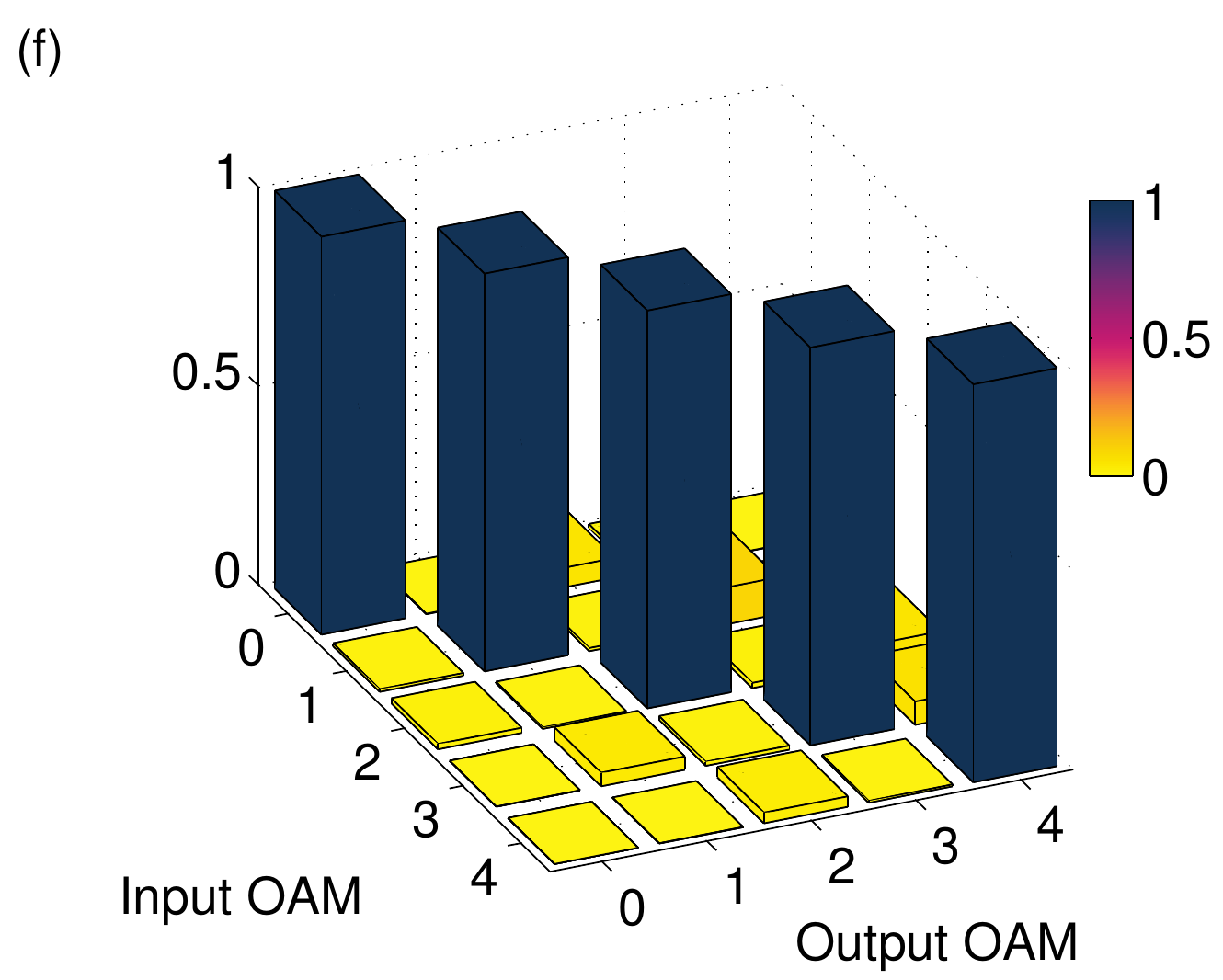}
\caption{(a-e) Streak camera images for OAM eigenstates ($\ell=0,2,4$) normalized based on~(\ref{eq:p_pbs}). (f) Tabulated results of the OAM spectrometer up to $\ell=4$.}
\label{fig:l0vppPatch}
\end{figure}

Finally, we benchmark the performance of the OAM spectrometer by comparing the correct versus incorrect detection of an input OAM eigenstate.  Figure~\ref{fig:l0vppPatch}(a-e) show examples of the output pulse sequences for input states $\ell_0=0-4$.  The integrated intensity under each peak versus the input state $\ell_0$ and the output time bin $\ell$ is shown in figure~\ref{fig:l0vppPatch}(f), where each row has been renormalized to the peak of the correct detection.  We define the crosstalk to be the ratios between the left and right adjacent incorrect detections versus the correct detection, i.e., $\frac{P(\ell_0=\ell,N=\ell\pm 1)}{P(\ell_0=\ell,N=\ell)}$ where $P(\ell_0,N)$ refers to the intensity measured at time $NT$ with input OAM state $\ell_0$.  For $\ell_0=0\mbox{ to }4$ we measured crosstalk values: $-12.3$, $-18.8$, $-20.0$, $-21.1$, $-21.6$, $-23.2$, $-24.8$ and $-28.2$ dB.  The geometric mean of the crosstalk is $7.47 \times 10^{-3}$ or $-21.3$~dB.

\section{Analysis of the spectrometer performance}

Below we simulate the performance of the spectrometer and analyze the main sources of error.  We model the laser pulses as a superposition of Laguerre-Gaussian (LG) modes, which are a complete orthonormal set of solutions to the paraxial wave equation.  Each optical element operating on the pulse maps each LG mode into a different superposition of other LG modes.  In particular, we focus on SMF, VPP and SLM.  The remaining optics: mirrors, beam splitters, waveplates, etc. are insensitive to the transverse mode of the beam and thus preserve the LG mode.  We then simulate the propagation of an input pulse through the spectrometer, and the coefficients of superposition obtained after the SMF corresponds to the measurement results.

\subsection{Laguerre-Gaussian modes}

The electric fields of LG modes can be described  in cylindrical coordinates $(\rho, \phi, z)$ as \cite{yao_orbital_2011}: 

        \begin{eqnarray}
        \fl
        u_{\ell,p}(\rho,\phi,z) = \langle\rho,\phi,z|\ell,p\rangle = \frac{1}{w(z)} \sqrt{ \frac{2}{\pi} \frac{p!}{(p+|\ell|)!}} \left(\frac{\sqrt{2}\rho}{w(z)}\right)^{|\ell|} \exp\left({-\frac{\rho^2}{w^2(z)}}\right)
        L_{p}^{|\ell|}\left(\frac{2\rho^2}{w^2(z)}\right)  \rme^{\rmi \ell\phi} \nonumber\\
        \exp\left(\frac{\rmi k_0 \rho^2 z}{2\left(z^2+z_{\rm R}^2\right)}\right) \exp\left({-\rmi(2p + |\ell| + 1)\tan^{-1}\left(\frac{z}{z_{\rm R}}\right)}\right)
        \label{eq:LG}
      \end{eqnarray}

Here $\ell\hbar$ ($\ell \in \mathbb{Z}$) is the OAM per photon, $p\geq 0$ labels the radial modes, $w(z)$ is the beam waist, $z_{\rm R}$ is Rayleigh range, and $k_0$ is the wave number of the fundamental Gaussian mode.

\subsection{Single mode fibre}
\label{sec:SMF_misalignment}

Ideally, an SMF selects only the fundamental Gaussian mode ($\ell=0,p=0$) while all the other orthogonal spatial modes do not propagate through.  In reality, higher order LG modes may still couple through the fibre due to finite apertures of optics and the fibre, imperfections of the fibre, mismatched beam waists between free-space and fibre-modes, and transverse misalignment between the propagation axes of the free-space and fibre-modes.  In our experiments, the last effect, the transverse misalignment, is by far the dominant, while the other effects were too small to be measured.  Hence we restrict our discussion to the transverse misalignment only.  For a misalignment of $\Delta$ (in units of beam waist) between the two optical axes, the coupling efficiency of the LG beam $(\ell,p)$ through the SMF is given by~(\ref{sec:app_gaussian_lg_overlap}):

\begin{equation}
      |{}_\Delta\langle 0,0|\ell,p\rangle|^2 = \frac{1}{p!(p+|\ell|)!} \left(\frac{\Delta^2}{2}\right)^{2p+|\ell|} \rme^{-\Delta^2}
\label{eq:type2misalignment}
\end{equation}

Experimentally, the above misalignment of the SMF can be estimated using the data for calibrating the energy distribution, as shown in figure~\ref{fig:results}(a). Nonzero output intensity from an input pulse of $\ell_0=1$ modes implies a misaligned (or imperfect) SMF. The amount of misalignment $\Delta$ can be calculated from the ratio of the two output intensities corresponding to the two input modes $\ell_0=0$ versus $\ell_0=1$ via:
\begin{equation}
\frac{P(\ell_0=1)}{P(\ell_0=0)} = \frac{|{}_\Delta\langle 0,0|M_{\mbox{VPP1}}|0,0\rangle|^2}{|{}_\Delta\langle 0,0|0,0\rangle|^2}
\label{eq:type2ratio}
\end{equation}
\begin{equation}
|{}_\Delta\langle 0,0|M_{\mbox{VPP1}}|0,0\rangle|^2 = \frac{\pi}{8}\Delta^2 \rme^{-\frac{3}{2}\Delta^2}\left(I_0\left(\frac{\Delta^2}{4}\right)+I_1\left(\frac{\Delta^2}{4}\right)\right)^2
\label{eq:type2misalignmentExact}
\end{equation}

The left hand side of~(\ref{eq:type2ratio}) is measured experimentally. $M_{\mbox{VPP1}}$ represents the operation by the SLM to create the input beam with $\ell_0=1$ and in a superposition of different $p$-states \cite{karimi_hypergeometric-gaussian_2007}.  Equation~(\ref{eq:type2misalignmentExact}) is obtained using~(\ref{eq:type2misalignment}) and~(\ref{eq:overlap}) (See~\ref{sec:app_gaussian_vpp_overlap} for details).  We plotted the derived misalignments in figure~\ref{fig:results}(b).  The misalignment is less than~10.1\% for all loops and converges to~3.7\% for higher loop numbers, indicating only a minor accumulative loop misalignment.

\subsection{VPP/SLM}
The VPP and SLM are used to change the OAM state.  The SLM, when applying the forked phase pattern, is mathematically equivalent to VPP for small diffraction angles \cite{heckenberg_laser_1992,bekshaev_spatial_2008}.  For a VPP of topological charge $\beta$, its operation on the laser beam can be described by a four-dimensional tensor $M_{VPP\beta}(z)$.  We solve for the tensor elements $m_{\ell_1,p_1;\ell_2,p_2;\beta}(z)$ analytically~(\ref{sec:app_vpp_tensor}) in the LG-mode basis:
      \begin{eqnarray}
      \fl
        m_{\ell_1,p_1;\ell_2,p_2;\beta}(z)  =  \left[\sqrt{\frac{p_1! p_2!}{(p_1+|\ell_1|)! (p_2 + |\ell_2|)!}} \right. 
  \sum_{k=0}^{p_1} \sum_{m=0}^{p_2} (-1)^{k+m}
 \frac{ (p_1 + |\ell_1|)! }{ (p_1 - k)! (|\ell_1| + k)! k!} 
  \nonumber \\
 \frac{ (p_2 + |\ell_2|)! }{ (p_2 - m)! (|\ell_2| + m)! m!} 
 \left. \Gamma\left( \frac{|\ell_1|+|\ell_2|}{2} + k + m + 1\right)  \right] 
 \nonumber \\
 \left[\exp\left( \rmi (2(p_1-p_2) + |\ell_1| - |\ell_2|)\tan^{-1}\left(\frac{z}{z_{\rm R}}\right)\right)\right]
 \nonumber \\
 \left[\frac{ \exp[2\pi \rmi(\ell_2 + \beta - \ell_1)] - 1}{2\pi \rmi (\ell_2 + \beta - \ell_1)}\right]
 \label{eq:overlap}
\end{eqnarray}

This equation consists of a product of three terms denoted by square brackets.  The first term with the double sum comes from the amplitude overlap between the different LG modes.  The second term, consisting of an exponential, describes the effects of Gouy phases.  The third term comes from OAM conservation and can describe the error in the topological charge of the VPP.  We discuss the effect of each term below.

\subsubsection{Gouy phase effects}

Different LG modes have different Gouy phases which also vary differently as the beam propagates; therefore, whenever there is a superposition of modes, there can be interference effects between the modes that varies under free space propagation \cite{arlt_generation_2000} and can lead to the loss of efficiency.  However, the effect does not produce crosstalk between different $\ell$-states, as OAM remain unchanged.  In our experiment, the effect of the Gouy phase accounts for less than $0.05\%$ loss in efficiency. This upper bound of $0.05\%$ was calculated based on distances between the optics in the experimental setup and using~(\ref{eq:overlap}).  In general, the effect is negligible when the propagation distance is much smaller than the Rayleigh range ($z \ll z_{\rm R}$).  When the size of the OAM spectrometer becomes comparable to the Rayleigh range, the use of~4-f systems between all phase elements would eliminate any Gouy phase effects.

\subsubsection{Topological charge error}

If the topological charge of the VPP is an integer, then the OAM of the beam is changed by that amount.  If the laser's wavelength is different from the nominal wavelength of the VPP, or if the beam comes at a skew angle to the VPP, then the VPP will appear thicker or thinner.  As a result, the OAM of the beam will be changed by a fractional value instead, or equivalently, become a superposition of LG modes of many OAM values.  This leads to a loss in both the efficiency and fidelity of the OAM spectrometer.

Such topological charge error can be modelled by non-integer $\beta$ in~(\ref{eq:overlap}).  In our experiment, both the laser wavelength and the angle of incidence are tightly controlled.  Even if we consider a rather generous~0.5\% error in the topological charge, it results in less than~0.05\% loss of efficiency and less than~-34dB~crosstalk.  Therefore the error in the topological charge is negligible.

\subsubsection{Lateral misalignment}

Although topological charge error will reduce the fidelity, it is, by far, not the leading cause.  If the optical axes of the beam and VPP are displaced relative to each other, the VPP produces a superposition of not just $p$, but $\ell$ states \cite{molina-terriza_management_2001} and thus reduces the fidelity.  Such lateral misalignment of the VPP affects all three terms in~(\ref{eq:overlap}).  To model it, we first express the VPP tensor $M_\beta=\rme^{\rmi\beta\phi}$ in a coordinate displaced from the common optical axis of the spectrometer.  We then numerically evaluate $m_{\ell_1,p_1;\ell_2,p_2;\beta}(z)$ in a subspace of ${\ell,p}=[-7,7]\times [0,10]$.  This chosen subspace yields less than~2\% error on theoretical crosstalk.

To compare with the experimental result, we calculated the average crosstalk versus VPP misalignment, assuming a transverse misalignment of the SMF of~3.7\% to~10.1\% as obtained in~\ref{sec:SMF_misalignment}.  The results are shown in~\ref{fig:total_results}(a), where the x-direction corresponds to the direction of SMF misalignment.  The measured crosstalk of $-21.3$ dB (\ref{sec:results}) corresponds to a VPP misalignment of~4.0\% to~6.1\% of the beam waist $w_0$.  A larger (smaller) VPP misalignment is allowed when its direction is aligned with (perpendicular to) that of the SMF misalignment.

\subsection{Limiting factors of fidelity}
In short, we show that misalignments of~3.7\% to~10.1\% at the SMF and~4.0\% to~6.1\% at the VPP are the leading causes for the loss of fidelity in our experiment, with an average crosstalk of~-21.3~dB.  Gouy phase effects do not affect fidelity and topological charge error contributes less than~-34~dB to crosstalk.

\begin{figure}
\includegraphics[width=0.33\textwidth]{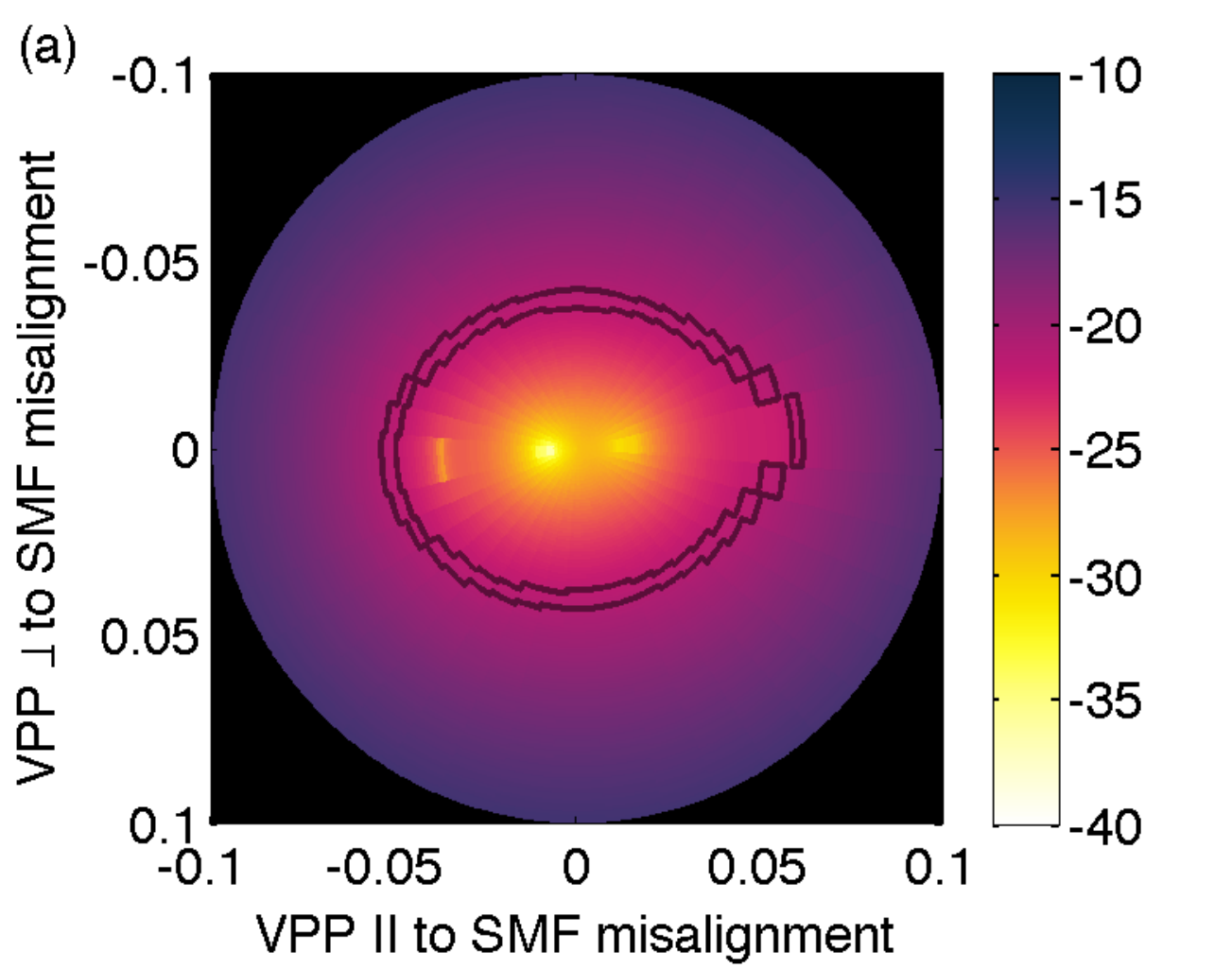}
\includegraphics[width=0.33\textwidth]{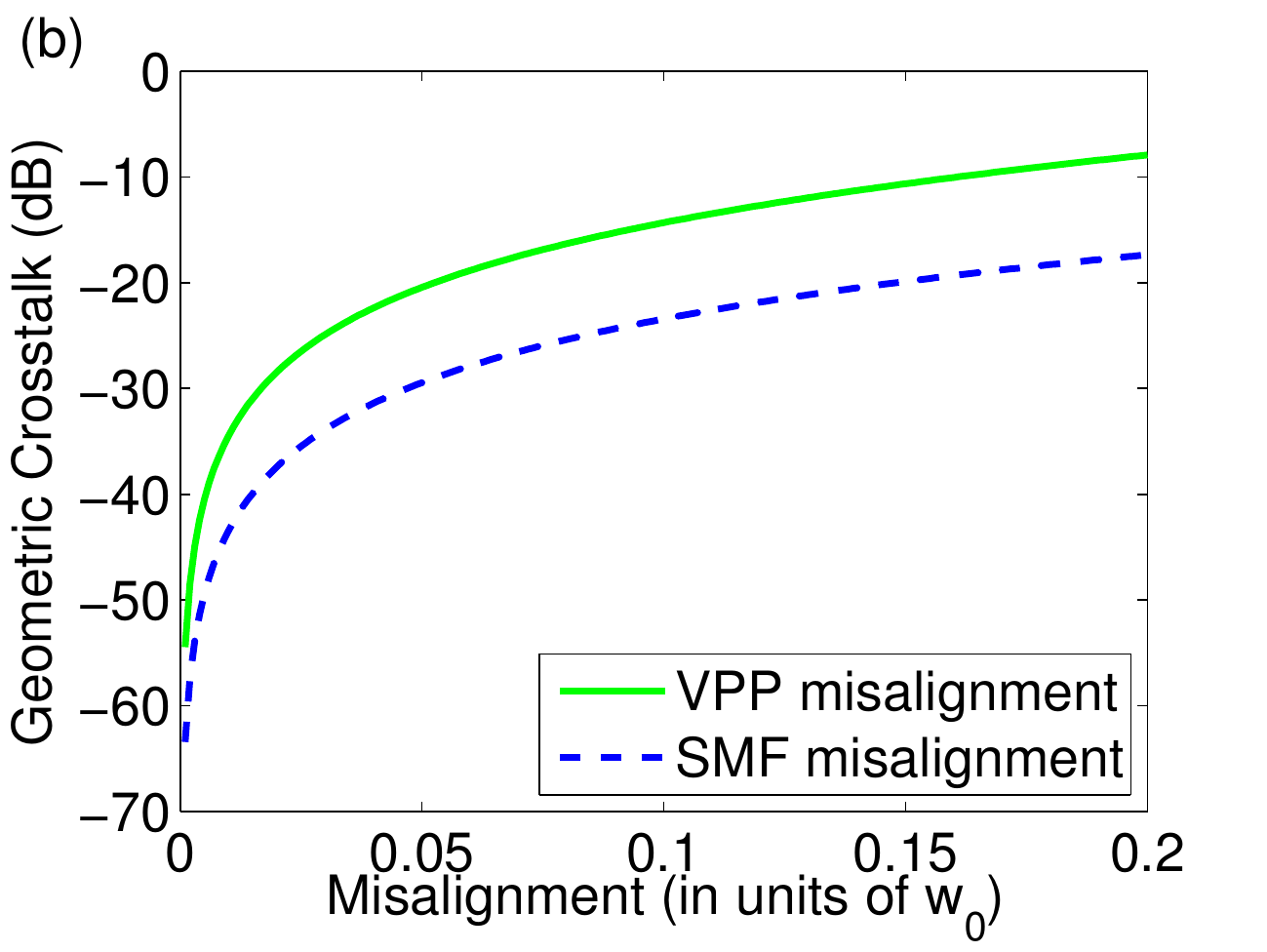}
\includegraphics[width=0.33\textwidth]{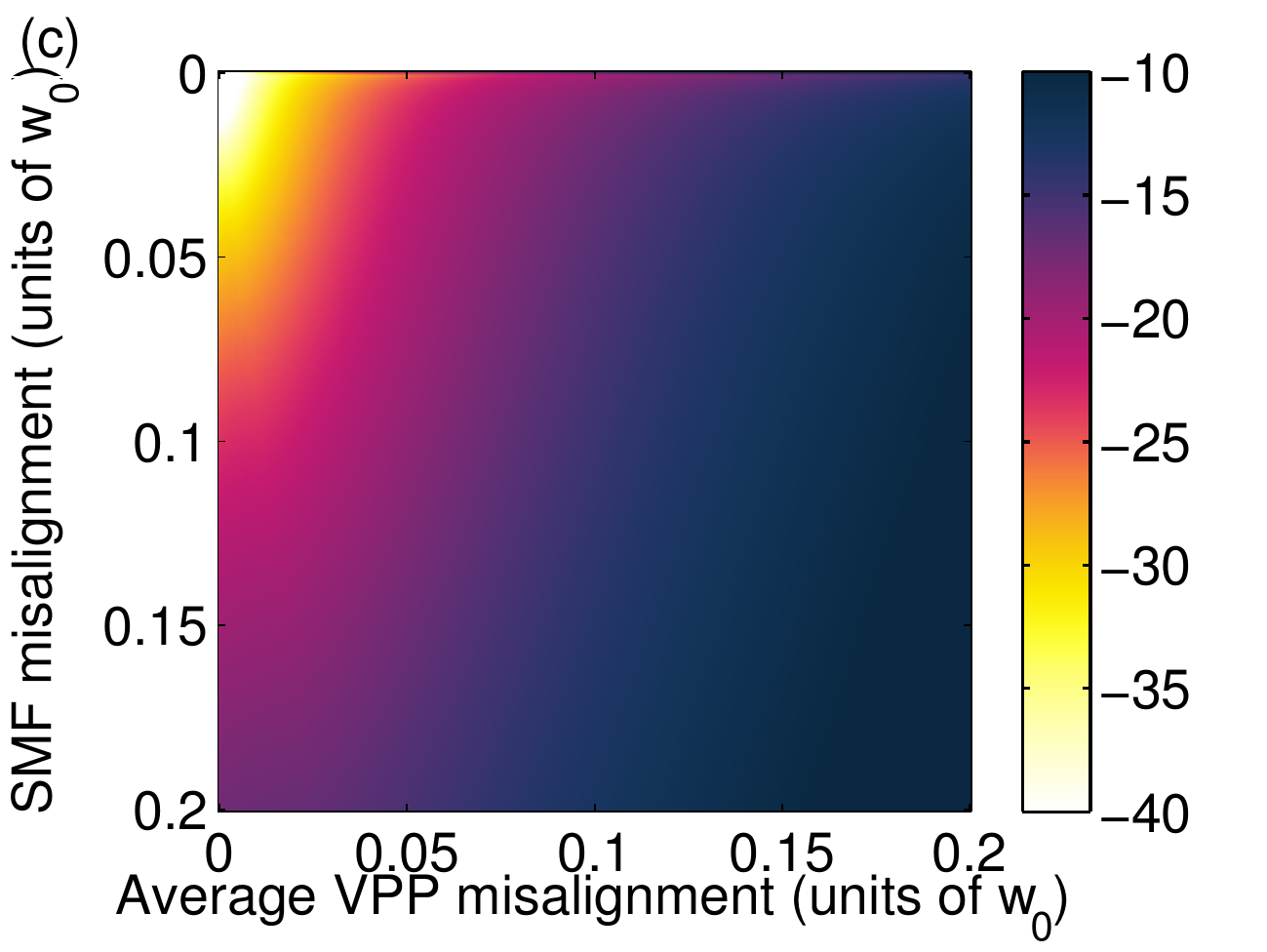}
\caption{Calculated the geometric mean of crosstalk (in dB) as a function of the lateral misalignment of the SMF and VPP (normalized by $w_0$).  (a) Crosstalk (in dB) as a function of the lateral misalignment of the VPP.  SMF misalignment (measured in~\ref{fig:results}(c)) is taken along the x direction for convenience, since only the relative angle between VPP and SMF misalignment is important.  The measured crosstalk of~$-20.9$ to~$-21.6$~dB corresponds to the values between the two black lines.  (b) The crosstalk versus VPP misalignment with no SMF misalignment (green solid line), and the crosstalk versus SMF misalignment with no VPP misalignment (blue dashed line).  (c) The upper bound of crosstalk (in dB) as a function of the SMF and VPP misalignments.  The upper bound corresponds to when the directions of the SMF and VPP misalignments are perpendicular to each other.  The crosstalk can be significantly reduced if the directions of the SMF and VPP lateral displacements are anti-parallel to each other.}
\label{fig:total_results}
\end{figure}

The fidelity of the spectrometer can be further increased by improving the optical alignment at the VPP and SMF.  We show in figure~\ref{fig:total_results}(b) the average crosstalk as a function of the VPP misalignment alone, neglecting SMF misalignment, and as a function of the SMF misalignment alone, neglecting VPP misalignment.  It is more sensitive to VPP misalignment than to SMF misalignment, but decreases superlinearly with the reduction of either misalignment.  With a VPP (SMF) misalignment of~$<1\%$, the crosstalk is reduced to~$<-34.4$~dB~($<-43.4$~dB).  When including both SMF and VPP misalignments, the crosstalk varies depending on the relative angle between the directions of the two misalignments.  The crosstalk is the greatest when the misalignments are in perpendicular directions from each other.  We show this upper bound crosstalk in~\ref{fig:total_results}(c).  When the misalignments are in opposite directions, we generally see a reduction in crosstalk, sometimes up to 20 dB.  Curiously, if the SMF is misaligned, crosstalk can be \emph{decreased} by introducing some VPP misalignment, and vice versa.  This is because with multiple elements misaligned, they can compensate each other.  However, zero crosstalk is reached only when both misalignments are exactly zero.

As misalignment is reduced, other subtle effects will need to be taken under consideration.  Such as imperfections in the surface roughness or other aberrations in the optics, which could further reduce efficiency and fidelity.  Eventually, the fidelity will be limited by overlap in time between adjacent pulses.  We did not include these effects in our model.

\section{Conclusions}
In summary, we present a practical OAM spectrometer that can map OAM to time up to arbitrarily large OAM values.  We have shown an average nearest neighbor crosstalk of~-21.3dB among~5 OAM states, limited mainly by optical misalignment.  The high fidelity of the demonstrated OAM spectrometer may enable accurate measurements of topological properties of objects such as in spiral imaging \cite{torner_digital_2005} and the angular momentum of black holes, which is encoded into the OAM spectrum of light from the accretion disc due to strong gravitational effects predicted by general relativity \cite{tamburini_twisting_2011}.  We also demonstrated speeds (80~MHz) orders of magnitude faster than the switching times of SLMs.  Miniaturization of optics could allow for GHz detection rate.

\section*{Acknowledgments}

We would like to thank Professor Miles Padgett for his helpful advice.  And also thank Professor Theo Lasser and Matthias Geissb\"{u}hler for their \emph{Morgenstemning} colormap \cite{geissbuehler_how_2013}.

\appendix
\section{}
\label{sec:app_lg_overlap}

In this appendix section we seek to derive the various actions and overlaps between the LG modes and optics.  We first derive the action of the VPP in the LG basis.  Then we derive the overlap of any LG mode with a misaligned SMF.  Lastly, we derive the overlap of OAM states created by a VPP with a misaligned SMF.  This is done by explicit integration and definitions of various special functions.  We use the inner product between Laguerre-Gaussian modes defined in~\ref{eq:LG} in cylindrical coordinates:

\begin{eqnarray}
\label{eq:lg_inner_product}
\fl\langle \ell_1,p_1|\ell_2,p_2\rangle = \int_0^\infty \rho \rmd\rho \int_0^{2\pi}  \rmd\phi \langle\ell_1,p_1|\rho,\phi,z\rangle \langle \rho,\phi,z|\ell_2,p_2\rangle
\nonumber \\
\times \int_0^\infty \rho \rmd\rho \int_0^{2\pi}  \rmd\phi u_{\ell_1,p_1}^*(\rho,\phi,z) u_{\ell_2,p_2}(\rho,\phi,z)
\end{eqnarray}

\subsection{Derivation of VPP tensor}
\label{sec:app_vpp_tensor}

We derive the four dimensional tensor of the action of a VPP in the LG basis as seen in~(\ref{eq:overlap}).  We start by explicitly writing down the overlap integral:

\begin{equation}
\fl\langle \ell_1,p_1| M_{\mbox{VPP}\beta} | \ell_2,p_2\rangle = \int_0^\infty \rho\rmd\rho \int_0^{2\pi} \rmd\phi LG^*_{\ell_1,p_1}(\rho,\phi) \rme^{+\rmi \beta\phi} LG_{\ell_2,p_2}(\rho,\phi)
\label{eq:vpp-overlap1}
\end{equation}

The LG functions are given by~(\ref{eq:LG}).  Due to separability of variables, we will solve the $\phi$-integral first:

\begin{equation}
\int_0^{2\pi} \rmd\phi \rme^{-\rmi \ell_1 \phi} \rme^{+\rmi \beta\phi} \rme^{+\rmi \ell_2 \phi} = \frac{ \exp\left(2\pi \rmi\left( \ell_2 + \beta - \ell_1\right)\right) - 1}{\rmi\left(\ell_2 + \beta - \ell_1\right)}
\label{eq:vpp-overlap2}
\end{equation}

In the limit where $\beta$ is an integer, the $\phi$-integral yields the OAM conserving solution of $2\pi \delta_{\ell_1 + \beta - \ell_2,0}$, where $\delta_{a,b}$ is the Kronecker delta.  The remaining $\rho$-integral is found by expanding the rest of the LG modes and the generalized Laguerre polynomials given by~(\ref{eq:laguerre}).  This yields a finite polynomial of degree $|\ell_1| + |\ell_2| + 2p_1 + 2p_2 + 1$ (the +1 is from the Jacobian) multiplied by the Gaussian function $\exp\left(\frac{-\rho^2}{2 w_0^2}\right)$.  This integral can be easily solved by using one of the definitions of the Gamma function as shown in~(\ref{eq:gamma_integral}) to yield the result in the text~(\ref{eq:overlap}).

\begin{equation}
L_p^k(x) = \sum_{m=0}^p (-1)^m \frac{ (p+k)! }{ (p-m)!(k+m)! m! } x^m
\label{eq:laguerre}
\end{equation}

\begin{equation}
\int_0^\infty \rmd x x^k \rme^{-x^2} = \frac{1}{2}\Gamma\left(\frac{1+k}{2}\right)
\label{eq:gamma_integral}
\end{equation}

\subsection{Overlap between fundamental Gaussian and Laguerre-Gaussian}
\label{sec:app_gaussian_lg_overlap}
We first derive the overlap between a misaligned fundamental Gaussian mode and generic higher order LG modes.  This is~(\ref{eq:type2misalignment}) in the text.  We start with the fundamental Gaussian mode defined in cylindrical coordinates ($\rho, \phi, z$) with an origin shifted by $w_0\Delta$ along the $\phi=0$ direction.  The propagation direction is along the z-axis.

\begin{equation}
{}_\Delta\langle 0,0|\rho,\phi,z\rangle = \frac{1}{w_0}\sqrt{\frac{2}{\pi}} \exp\left({\frac{-\left(\rho^2 + 2\rho w_0\Delta \cos\phi + w_0^2 \Delta^2\right)}{w_0^2}}\right)
\label{eq:gaussian_overlap1}
\end{equation}

If the origin were shifted along a different $\phi$ angle, the only change would be the overall phase which does not affect the intensity.  These calculations are performed at the beam waist since the beam should be focused on the fibre to have any substantial coupling.  With a change of variables $x=\frac{\sqrt{2}\rho}{w_0}$ the complete overlap integral is now:
	  
\begin{eqnarray}
\label{eq:gaussian_overlap2}
\fl {}_\Delta \langle 0,0|\ell,p\rangle = \int_0^\infty \int_0^{2\pi} x\rmd x \rmd\phi \frac{1}{\pi} \sqrt{ \frac{p!}{(p+|\ell|)!}} x^{|\ell|} \exp\left(\frac{-x^2}{2}\right) L_{p}^{|\ell|}\left(x^2\right) \rme^{\rmi\ell \phi} 
\nonumber \\
\exp\left(-\left(\frac{x^2}{2} + \sqrt{2} x \Delta \cos\phi + \Delta^2\right)\right)
\end{eqnarray}

The $\phi$-integral is related to the well known Bessel integral~(\ref{eq:bessel_int}), so~(\ref{eq:gaussian_overlap2}) becomes~(\ref{eq:gaussian_overlap3}).

\begin{equation}
J_n(z) = \frac{1}{2\pi \rmi^n} \int_0^{2\pi} \rmd\phi \rme^{\rmi z\cos\phi} \rme^{\rmi n\phi} 
\label{eq:bessel_int}
\end{equation}

\begin{equation}
\fl {}_\Delta \langle 0,0|\ell,p\rangle =       \sqrt{ \frac{p!}{(p+|\ell|)!}} \int_0^\infty x \rmd x    x^{|\ell|} \rme^{-\left(x^2 + \Delta^2\right)} L_{p}^{|\ell|}\left(x^2\right) 2 \rmi^{-|\ell|} (-1)^{|\ell|} J_{|\ell|}\left(\rmi \sqrt{2} \Delta x\right)
\label{eq:gaussian_overlap3}
\end{equation}

The Bessel function can be converted into an infinite sum of generalized Laguerre polynomials as given in~(\ref{eq:bessel_laguerre}) below with $\sqrt{w} = \rmi\Delta / \sqrt{2}$.  Using~(\ref{eq:bessel_laguerre}) and the orthogonality relationship between generalized Laguerre polynomials~(\ref{eq:laguerre_weight}).  Equation~(\ref{eq:gaussian_overlap3}) magnitude squared surprisingly reduces to the rather simple expression of~(\ref{eq:type2misalignment}) in the text.

\begin{eqnarray}
\sum_{n=0}^\infty \frac{ L_n^k(x)}{\Gamma(n+k+1)} w^n = e^w (xw)^{\frac{-k}{2}} J_k\left(2\sqrt{xw}\right)
\label{eq:bessel_laguerre}
\end{eqnarray}

\begin{eqnarray}
\int_0^\infty \rmd x \rme^{-x} L_n^k(x) L_m^k(x) = \frac{(n+k)!}{n!} \delta_{m,n}
\label{eq:laguerre_weight}
\end{eqnarray}

\subsection{Overlap between fundamental Gaussian and vortex phase plate mode}
\label{sec:app_gaussian_vpp_overlap}
We create states with OAM by passing a Gaussian beam through a VPP.   This state can be expressed as a sum of Laguerre Gaussian modes as given in~(\ref{eq:overlap_gaussian}), which is a special case of the more general equation solved in section~\ref{sec:app_vpp_tensor}.  In the previous section, we calculated the overlap of a shifted fundamental Gaussian mode and any Laguerre Gaussian mode.  Therefore, by combining these two calculations, we can derive the overlap of a shifted Gaussian mode and an OAM mode created by a VPP~(\ref{eq:type2misalignment_sum}), here $z = \frac{\Delta^2}{2}$.

\begin{eqnarray}
\label{eq:overlap_gaussian}
M_{\mbox{VPPN}}|0,0\rangle = \sum_{p=0}^\infty m_{0,0;N,p;N} |N,p\rangle \\ \nonumber
\mbox{where }m_{0,0;N,p;N} = \frac{N}{2} \sqrt{\frac{1}{(p+N)!}}\Gamma\left(p+\frac{N}{2}\right)
\end{eqnarray}

\begin{eqnarray}
\label{eq:type2misalignment_sum}
{}_\Delta\langle 0,0|M_{\mbox{VPPN}}|0,0\rangle = 
 \frac{e^{-z}z^{\frac{N}{2}} (-1)^{N} N}{2} \sum_{p=0}^{\infty} \frac{ (-z)^p \Gamma\left(p+\frac{N}{2}\right)}{(p+N)!p!}
\end{eqnarray}

Equation~(\ref{eq:type2misalignment_sum}) can be written in the form of a regularized confluent hypergeometric function~(\ref{eq:hypergeometric_1f1_form}) which can be evaluated using~(\ref{eq:reghypergeometric_1_1_a_2a-1_z})\footnote{\url{http://functions.wolfram.com/07.21.03.0011.01} } to yield the result in~(\ref{eq:type2misalignmentExactGeneral}), where $I_\alpha(x)$ is the modified Bessel function of the first kind.  The expression can be further simplified by noting $I_\alpha(-z) = \rmi^{2\alpha} I_\alpha(z)$.  In the case of $N=1$ and taking the magnitude squared, this reduces to the form in the text~(\ref{eq:type2misalignmentExact}).

\begin{equation}
{}_1\tilde{F}_1(a;b;z) = \sum_{k=0}^\infty \frac{\Gamma(a+k)}{\Gamma(b+k) \Gamma(a)} \frac{z^k}{k!}
\label{eq:hypergeometric_1f1_form}
\end{equation}

\begin{equation}
{}_1\tilde{F}_1\left(\frac{N}{2};N+1;z\right) = \frac{e^{\frac{z}{2}} \sqrt{\pi}}{2 \Gamma\left(\frac{N}{2}+1\right)} z^{\frac{1}{2}-\frac{N}{2}} \left( I_{\frac{N}{2}-\frac{1}{2}}\left(\frac{z}{2}\right) - I_{\frac{N}{2}+\frac{1}{2}}\left(\frac{z}{2}\right)\right)
\label{eq:reghypergeometric_1_1_a_2a-1_z}
\end{equation}

\begin{equation}
\frac{\rme^{\frac{-3\Delta^2}{4}} \Delta^2 \sqrt{\pi}}{2\sqrt{2}}
\left( I_{\frac{N}{2}-\frac{1}{2}}\left(\frac{\Delta^2}{4}\right) + I_{\frac{N}{2}+\frac{1}{2}}\left(\frac{\Delta^2}{4}\right)\right)
\label{eq:type2misalignmentExactGeneral}
\end{equation}

\section*{References}
\bibliographystyle{unsrturl}
\bibliography{simple_oam_sorter_reduce}

\end{document}